\makeatletter
\@namedef{ver@picins.sty}{9999/99/99}
\makeatother
\documentclass{SCIS2019}
\usepackage{bbm}
\usepackage{enumitem}
\usepackage{diagbox}
\usepackage{makecell}
\usepackage{tabularx,multirow}
\begin{document}
\ArticleType{RESEARCH PAPER}
\Year{2019}
\Month{}
\Vol{}
\No{}
\DOI{}
\ArtNo{}
\ReceiveDate{}
\ReviseDate{}
\AcceptDate{}
\OnlineDate{}

\title{Decoding Chinese phonemes from intracortical brain signals with hyperbolic-space neural representations}{Decoding Chinese phonemes from intracortical brain signals with hyperbolic-space neural representations}

\author[1,2,3]{Xianhan Tan}{}
\author[4]{Junming Zhu}{}
\author[4]{Jianmin Zhang}{}
\author[1,3]{Yueming Wang}{}
\author[3,5,6]{Yu Qi}{{qiyu@zju.edu.cn}}

\AuthorMark{Xianhan Tan}

\AuthorCitation{Xianhan Tan, et al.}


\address[1]{Qiushi Academy for Advanced Studies, Zhejiang University, Hangzhou {\rm 310027}, China}
\address[2]{College of Computer Science, Zhejiang University, Hangzhou {\rm 310027}, China}
\address[3]{State Key Lab of Brain-Machine Intelligence, Hangzhou {\rm 310027}, China}
\address[4]{Second Affiliated Hospital of Zhejiang University School of Medicine, Hangzhou {\rm 310027}, China}
\address[5]{Affiliated Mental Health Center and Hangzhou Seventh People's Hospital, Zhejiang University School of Medicine, Hangzhou {\rm 310027}, China}
\address[6]{MOE Frontier Science Center for Brain Science and Brain-machine Integration, Zhejiang University, Hangzhou {\rm 310027}, China}


\abstract{Speech brain-computer interfaces (BCIs), which translate brain signals into spoken words or sentences, have shown significant potential for high-performance BCI communication. Phonemes are the fundamental units of pronunciation in most languages.  While existing speech BCIs have largely focused on English, where words contain diverse compositions of phonemes, Chinese Mandarin is a monosyllabic language, with words typically consisting of a consonant and a vowel. This feature makes it feasible to develop high-performance Mandarin speech BCIs by decoding phonemes directly from neural signals. This study aimed to decode spoken Mandarin phonemes using intracortical neural signals. We observed that phonemes with similar pronunciations were often represented by inseparable neural patterns, leading to confusion in phoneme decoding. This finding suggests that the neural representation of spoken phonemes has a hierarchical structure. To account for this, we proposed learning the neural representation of phoneme pronunciation in a hyperbolic space, where the hierarchical structure could be more naturally optimized. Experiments with intracortical neural signals from a Chinese participant showed that the proposed model learned discriminative and interpretable hierarchical phoneme representations from neural signals, significantly improving Chinese phoneme decoding performance and achieving state-of-the-art. The findings demonstrate the feasibility of constructing high-performance Chinese speech BCIs based on phoneme decoding.
}



\maketitle

\section{Introduction}

Speech is a remarkable ability unique to humans, allowing for precise and effective communication. Speech brain-computer interfaces (BCIs), which directly translate brain signals into spoken words or sentences, offer tremendous potential to establish an ideal communication pathway for individuals with aphasia. Recent years have seen significant advancements in speech BCIs \cite{moses2021neuroprosthesis,stavisky2018decoding,stavisky2019neural,wilson2020decoding}, enabling direct speech synthesis and decoding of spoken words and sentences from neural signals recorded from the cerebral cortex \cite{bouchard2013functional,lotte2015electrocorticographic, moses2019real}.

Existing speech BCIs have predominantly focused on the English language, with the aim of producing speech acoustics \cite{moses2019real} or classifying spoken words and sentences \cite{moses2021neuroprosthesis} from neural signals. In a recent study \cite{moses2021neuroprosthesis}, a speech BCI system assisted a patient with aphasia in communicating through a dialogue system by classifying 50 words and sentences online, achieving a median word error rate of 25.6\% and demonstrating the feasibility of speech BCIs for clinical applications. However, directly decoding spoken words from neural signals faces the critical challenge of limited vocabulary size, as the subject must repeatedly read the words in the vocabulary for decoder training, which can be highly time-consuming.

Phonemes are the fundamental units of sound in language, and their number is typically much smaller than the number of words. Accurately recognizing phonemes can enable general speech decoding. However, in English, which has numerous phonemes per word, the assimilation and linking of sounds in pronunciation make phoneme-based decoding challenging \cite{stavisky2018decoding, moses2016neural}. Prior studies have attempted to classify phonemes in English using neural signals, with \cite{stavisky2018decoding} achieving an average accuracy of 33.9\% for 39 phonemes and \cite{mugler2014direct} achieving an average accuracy of 36.1\% for 24 consonants.

In contrast to English, Mandarin Chinese is a monosyllabic language where words typically consist of a consonant and a vowel in an `initial-final' structure. This unique feature makes it possible to develop high-performance Mandarin speech BCIs by directly decoding phonemes from neural signals. In this study, we aimed to decode spoken Mandarin phonemes using intracortical neural signals (single-unit activity). However, we observed that phonemes with similar pronunciations often shared inseparable neural patterns, leading to confusion in phoneme decoding. This observation suggests that the neural representation of spoken phonemes in Mandarin has a hierarchical structure, which has also been reported in speech recognition and neuroscience studies \cite{kimball1992context, odell1994tree,mesgarani2014phonetic, moses2016neural}.

To address this issue, we propose a novel approach to learning the neural representation of phoneme pronunciation in a hyperbolic space \cite{nickel2017poincare}. In this space, the capacity increases exponentially with the radius \cite{khrulkov2020hyperbolic, chami2019hyperbolic}, which is consistent with the hierarchical structure of phonemes. This allows for a more natural optimization of the hierarchical structure. To decode spoken phonemes from neural signals, we introduce a hyperbolic neural network. Although studies suggest the existence of hierarchical structures in neural representations of phonemes, the specific structure is unknown. Therefore, we construct a latent hierarchical constraint by gradient-based hyperbolic hierarchical clustering to encourage the neural representations to learn an optimal hierarchical structure with a hierarchical clustering process which is jointly optimized with the phoneme classification objective in the network training.
The main contributions of this study are summarized as follows:
\begin{itemize}
\item We find that the neural representation of spoken phonemes contains a hierarchical structure, which is mostly shared with the articulation structures of spoken phonemes.
\item We propose a hyperbolic neural network to learn effective neural representations of Mandarin phonemes so that phonemes with similar pronunciations can be clearly separated.
\item Experiments with clinical neural signals from a Mandarin-speaking human participant demonstrate that the use of hyperbolic metrics leads to a more efficient representation of phonemic neural signals.
\end{itemize}

The hyperbolic space-based approach provides a novel perspective for neural representation learning and neural decoding. The findings suggest the feasibility of constructing high-performance Chinese speech BCIs based on decoding Mandarin consonants and vowels.

\section{Clinical Experiment and Neural Signal Acquisition}

We recorded neural activity from a Chinese Mandarin-speaking participant in this study \cite{qi2022dynamic}. The participant was implanted with two 96-channel Utah intracortical microelectrode arrays (Blackrock Microsystems, Salt Lake City, UT, USA) in the left primary motor cortex, with one array positioned in the middle of the hand knob area (array-A) and the other located medially about 2mm apart (array-B). All clinical and experimental procedures were approved by the Medical Ethics Committee of The Second Affiliated Hospital of Zhejiang University (Ethical review number 2019-158, approved on 05/22/2019) and were registered in the Chinese Clinical Trial Registry (Ref. ChiCTR2100050705).

\begin{figure*}[t]
  \centering
  \includegraphics[width=1\linewidth]{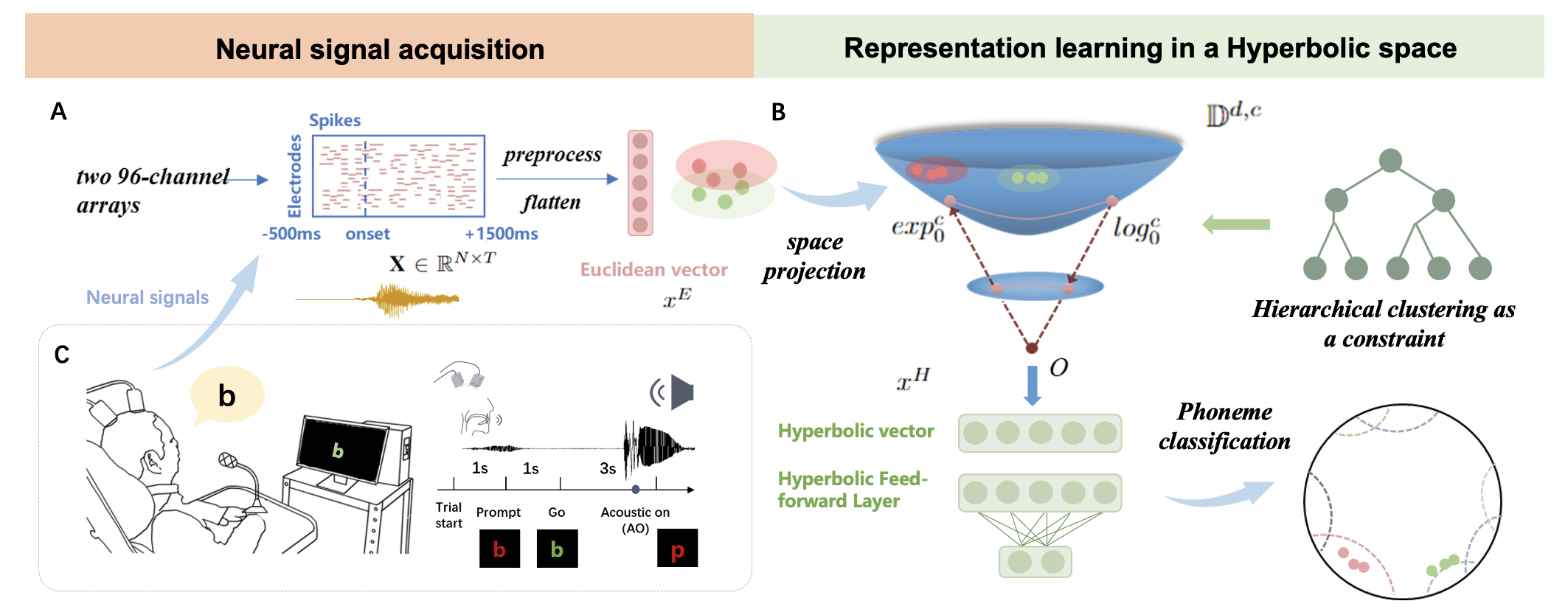}
  \caption{The flowchart of HYSpeech-based Mandarin phoneme decoding. (A) Neural signal recording and preprocessing. (B) Projection of neural signals into the hyperbolic space. A clustering loss and a classification loss are jointly optimized to learn neural representations with hierarchical structures. (C) Illustration of the experimental paradigm.}
  \label{fig:model}
\end{figure*}

The speech task designed for the participant required him to say a Mandarin phoneme or syllable on each trial, as illustrated in Figure \ref{fig:model}C. Mandarin is a monosyllabic language, and each Mandarin syllable consists of a Mandarin consonant and a Mandarin vowel, with a single Mandarin syllable being treated as a word. For clarity, all phonemes and words mentioned below are Mandarin-specific.
\begin{itemize}
\item
\textbf{Experimental paradigm:} The experiments consisted of three speaking tasks: 1) 21 different consonants, 2) 24 different vowels, and 3) 20 different words. In each trial, the participant was instructed to watch a red phoneme on a computer screen placed a meter in front of them and hear a vocal cue for the phoneme, as depicted in Figure \ref{fig:model}C. The phoneme turned green after one second, indicating the start of the `Go' stage where the participant was required to say the prompted phoneme. To ensure sufficient response time, the `Go' stage lasted for three seconds. After the `Go' stage, the trial ended, and the next trial commenced.
\item
\textbf{Neural signal acquisition:} Neural signals were acquired at 30 kHz using a 256-channel Neuroport system (NSP, Blackrock Microsystems) with two 96-channel Utah intracortical microelectrode arrays. During the experiment, the audio signals were recorded using a microphone placed in front of the participant at 30 kHz by the NeuroPort system via an analog input port.
\item
\textbf{Signal processing:} The raw spike signals were first sorted into single-unit activity using the Offline Sorter software. To identify the temporal segments where the phonemes were spoken, the audio signals were manually annotated using the Praat software package to annotate the start and end timestamps of the period when the participant spoke the phonemes as the `acoustic on (AO)' stage. For each trial, the data segment from 0.5 seconds before the `AO' stage to 1.5 seconds after it was used. The spike signals were binned into spike counts with a 100 ms time window and a stride of 25 ms. After processing, a trial was represented with a matrix $\textbf{X} \in \mathbb{R}^{N \times T}$, where $N$ denotes the number of neurons, and $T$ denotes the number of time bins. Finally, the matrix was flattened into a vector $x^E$ in Euclidean space.
\end{itemize}

The clinical dataset contains neural signals from four experimental days. Each day contains 3-4 sessions, and each session includes more than 20 Mandarin phonemes or words.

\section{Hyperbolic Network for Chinese Phoneme Decoding}
In this section, we present the proposed hyperbolic network model (HYSpeech) to decode spoken phonemes from neural signals. The framework of the proposed approach is illustrated in Figure \ref{fig:model}.

\subsection{Tree-based structure of Mandarin phonemes according to the articulations}

In Chinese Mandarin, the phonemes can be classified based on the articulatory movements involved \cite{duanmu2007phonology}. In Figure A1 (see Appendix A), we illustrate the movements of the articulators during the pronunciation of consonants and vowels.
As the articulation of phonemes involves a sequential combination of the lip, teeth, tongue, gum, and palate, there is a hierarchical structure inherent in phoneme pronunciation. Therefore, it is natural to assume that the neural signals representing the pronunciation process may also contain hierarchical structures.

To account for this assumption, we propose to transform the neural signals into a hyperbolic space \cite{nickel2017poincare}, where tree-like structures or hierarchies are more readily distinguishable than in Euclidean space \cite{nickel2018learning}. In the hyperbolic space, the capacity of the space increases exponentially with the radius \cite{khrulkov2020hyperbolic, chami2019hyperbolic}, which is consistent with the properties of a tree. As the number of nodes in a tree increases exponentially with its depth, the hyperbolic space can enhance the discriminative ability of similar phonemes.

\subsection{Projecting neural signals to the hyperbolic space}

Firstly, the neural signals $\textbf{X} \in \mathbb{R}^{N \times T}$ are projected to the hyperbolic space. Hyperbolic space is a space with negative constant curvature which can be described by many isometric models, e.g., Poincar$\acute{e}$ disk model, Lorentz model. Here, we use the Poincar$\acute{e}$ disk model $(\mathbb{D}_c^d,g^c)$, which is the most commonly used and well-modeled in tree-structured data. In the Poincar$\acute{e}$ model, hyperbolic space can be described by Poincar$\acute{e}$ disk $\mathbb{D}_c^d = \{x \in \mathbb{R}^d : \parallel x \parallel < 1\}$, and any point on the hyperbolic space can be projected onto the disk (shown in Figure \ref{fig:model}) which equipped with the Riemannian metric: $g^c = \lambda_x^2g^E$ where $\lambda_x = \frac{2}{1 - \parallel x \parallel^2}$ is the conformal factor, $c$ is the curvature and $g^E$ is the Euclidean metric.

On the Poincar$\acute{e}$ disk, the distance between two points $x, y \in \mathbb{D}_c^d$ is defined as
 \begin{equation}
d^c(x, y) = cosh^{-1}(1 + 2\frac{\parallel x-y \parallel^2}{(1-\parallel x \parallel^2)(1- \parallel y \parallel^2)} )
\end{equation}
where $d^c(,)$ denotes the hyperbolic distance between points increases exponentially when the points are close to the boundary, which is well suited to the increasing depth of the tree. Thus, the Poincar$\acute{e}$ disk model is well suited to modeling a tree.

After modeling the hyperbolic space, we need to define the operations in the hyperbolic space. In Euclidean space, everywhere is flat, which makes vector operations easy to complete. In contrast, hyperbolic space is curved, and even the simplest translation operations in hyperbolic space are very complex. The most common method is to transfer the data and operations to the tangent space, which preserves the local Euclidean nature, and then return to the hyperbolic space when the operations are completed. Therefore, we introduce the gyrovector space, which defines the mutual transformation from tangent space to hyperbolic space based on the Möbius transformation.

Suppose $x^E$ is a Euclidean vector of neural signals and $x^E$ lies in the tangent spaces $T_p\mathbb{D}_c^d$ at the origin $p = \textbf{0}$. $x^E$ can be projected to hyperbolic space by the exponential map $exp_{\textbf{0}}^c : T_\textbf{0}\mathbb{D}_c^d \rightarrow \mathbb{D}_c^d$ :
\begin{equation}
    exp_{\textbf{0}}^c(x^E) = tanh(\sqrt{c}\parallel x^E \parallel)\frac{x^E}{\sqrt{c}\parallel x^E \parallel} = x^H
\end{equation}
where $x^H$ is the hyperbolic vector. In contrast, $x^H$ can be projected to Euclidean space by the logarithmic map  $log_{\textbf{0}}^c : \mathbb{D}_c^d \rightarrow T_\textbf{0}\mathbb{D}_c^d $:
\begin{equation}
    log_{\textbf{0}}^c(x^H) = tanh^{-1}(\sqrt{c}\parallel x^H \parallel)\frac{x^H}{\sqrt{c}\parallel x^H \parallel} = x^E.
\end{equation}
After projecting neural signals to the hyperbolic space, we get the hyperbolic vector $x^H$, to be used as the input to the hyperbolic network.

\subsection{Phoneme classification in the hyperbolic space}
Previous studies have demonstrated that hyperbolic neural networks can extract features of hyperbolic vectors well and perform well on various tasks, e.g. classification, and generation. Here, we use hyperbolic feed-forward layers (FFNN) \cite{ganea2018hyperbolic} $f^{\bigotimes_c}(x) : \mathbb{D}_c^d \rightarrow \mathbb{D}_c^l$ to extract features from $x^H$ as follow:
\begin{equation}
    f^{\bigotimes_c}(x^H) =  exp_{\textbf{0}}^c(f(log_{\textbf{0}}^c(x^H))) = x^L
\end{equation}
where $f : \mathbb{R}^d \rightarrow \mathbb{R}^l$ is the Euclidean function, $x^L$ is the latent vector and $l$ is  dimension of $x^L$.

With the two transformations mentioned above, we can define the vector operation in hyperbolic space:
\begin{equation}
    x\bigoplus_{c} y = \frac{(1 + 2c<x,y> + c \parallel y \parallel^2)x + (1 - c \parallel x \parallel^2)y}{1 + 2c<x,y> + c^2 \parallel x \parallel^2 \parallel y \parallel^2}
\end{equation}
\begin{equation}
    r \bigotimes_c x = (1/\sqrt{c})tanh(rtanh^{-1}(\sqrt{c\parallel x \parallel}))\frac{x}{\parallel x \parallel})
\end{equation}
where $\bigoplus_c$ denotes the Möbius addition and $\bigotimes_c$ denotes the Möbius scalar multiplication following the formalism of M$\ddot{o}$bius gyrovector spaces\cite{ungar2001hyperbolic,ungar2008analytic,ungar2008gyrovector}. Based on these operations, hyperbolic neural networks can be naturally defined.

We then use hyperbolic multiclass logistic regression (MLR) \cite{ganea2018hyperbolic} to get the logit possibility of classification. Given $k$ classes, we can define the logit possibility of $x^L$ as:
\begin{equation}
\begin{split}
    p(y = k | & x^L)  \propto  exp(\frac{\lambda_{p_k}^c\parallel a_k \parallel}{\sqrt{c}}sinh^{-1}\frac{2\sqrt{c}<-p_k \bigoplus_c x, a_k>}{(1 - c\parallel -p_k \bigoplus_c x \parallel^2)\parallel a_k \parallel} )
\end{split}
\end{equation}
where $a_k \in \mathbb{D}_c^d $ and $p_k \in T_p\mathbb{D}_c^d$ are trainable parameters, which defines a set of hyperbolic hyperplane: $\tilde{\mathcal{H}}_{a,p}^c = \{x \in \mathbb{D}_c^d : <-p \bigoplus_c x, a> = 0\}$ and $\lambda_{p_k}^c$ denotes conform factor.

Finally, we get the classification loss by softmax function and cross-entropy loss function:
\begin{equation}
L_{cls} = \sum_{i}-y_ilog(p_i)
\end{equation}
where $y_i$ is one-hot label of data, and $p_i$ is softmax possibility.

\subsection{Hierarchical clustering in the hyperbolic spaces}
Since the hierarchical structure in neural signals is not clear, we do not have a direct hierarchical structure as prior knowledge to help guide the representation learning process. Therefore, we propose a latent constraint to learn a proper hierarchical structure in hyperbolic space. To this end, we employ hierarchical clustering (HC), which can help mine the underlying relationships of data, to construct a binary tree based on the pairwise similarity of the data. The learning of the hierarchical tree is based on a cost function proposed by Dasgupta \cite{dasgupta2016cost}, in which a good tree should have a small cost:
\begin{equation}
C_{(T; w)} = \sum_{i,j}w_{ij}| lvs(T[lca(i,j)]).
\end{equation}
It allows nodes with high similarity ($w_{ij}$) to be merged first. $lca(,)$ denotes the lowest common ancestor of two leaf nodes, and $lvs(,)$ denotes the leaves of $T[lca(i,j)]$, which is a discrete operation and cannot be continuously optimized.
Study \cite{wang2018improved} rewrote the Dasgupta's Cost by triplets of datapoints $i, j, k$:
\begin{equation}
 C_{(T; w)} = \sum_{i,j,k}(w_{ij} + w_{ik} + w_{jk} - w_{ijk}(T;w) + 2\sum_{i,j}w_{ij}
\end{equation}
$$ w_{ijk}(T;w) = w_{ij}\mathbbm{1}{\{i,j|k\}} + w_{ik}\mathbbm{1}{\{i,k|j\}} + w_{jk}\mathbbm{1}{\{j,k|i}\}$$
which simplifies the calculation of $lca(,)$, and the relation ${\{i,k|j\}}$ holds if $lca(i,j)$ is a descendant of  $lca(i,j,k)$. Based on this formulation, recent works relax it using a continuous notion of $lca(,)$, such as gHHC\cite{monath2019gradient}, HypHC\cite{chami2020trees}, which could be optimized with gradient-descent.

Suppose $\textbf{X}_p = \{p_1,...,p_b\}$ is a batch of logit vectors, and $p_i,p_j,p_k$ is the triples sampled from  $\textbf{X}_p$. Then we could define our clustering  loss function as follow:
\begin{equation}
\begin{split}
 L_{tree} = \sum_{i,j,k} (\sigma_\tau(d_{ij}) + \sigma_\tau(d_{ik}) + \sigma_\tau(d_{jk}) & - \sigma_{\tau_{_\textbf{o}}}(d_{ijk}) + 2\sum_{i,j}\sigma_\tau(d_{ij})
\end{split}
\end{equation}
\begin{equation}
\begin{split}
\sigma_{\tau_{_\textbf{o}}}(d_{ijk}) & = (\sigma_\tau(d_{ij}), \sigma_\tau(d_{ik}), \sigma_\tau(d_{jk})) \cdot \sigma_\tau(d_{\textbf{o}}(lca(i,j)), d_{\textbf{o}}(lca(i,k)), d_{\textbf{o}}(lca(j,k))^T
\end{split}
\end{equation}
where $\sigma_\tau(d_i) = e^{d_i/\tau}/\sum_j e^{d_j/\tau}$, which is the scaled softmax function. $d_{ij}$ is hyperbolic distance of two logits vectors ($p_i$ and $p_j$), $d_{\textbf{o}}$ is hyperbolic distance to hyperbolic origin.


The loss function is specifically designed in two ways: (1) we use hyperbolic distance as the similarity metric, and (2) we calculate the similarity on the logits vectors. Thus, we can naturally combine our classification loss.
\begin{equation}
L_{joint} = \lambda L_{cls} + \gamma L_{tree}
\end{equation}
where  $\lambda$ and $\gamma$ are two hyper-parameters. We then apply the Riemannian stochastic gradient descent (RSGD) \cite{bonnabel2013stochastic} method to update the network parameters.

\section{Result and Analysis}
In this section, we first show the representation of neural signals recorded from our clinical experiments. Secondly, We analyze the phonemic similarities of neural activity based on the category of phonemes according to the articulations. The category of phonemes is given in Appendix A. We then analyze the phonemic structure learned by our model and compared the classification performance with other methods. Finally, we evaluate our model in different settings.

\subsection{Neural representation of spoken phonemes}

\subsubsection{Analysis of neural activities}
\begin{figure}[t]
  \centering
  \includegraphics[width=1\linewidth]{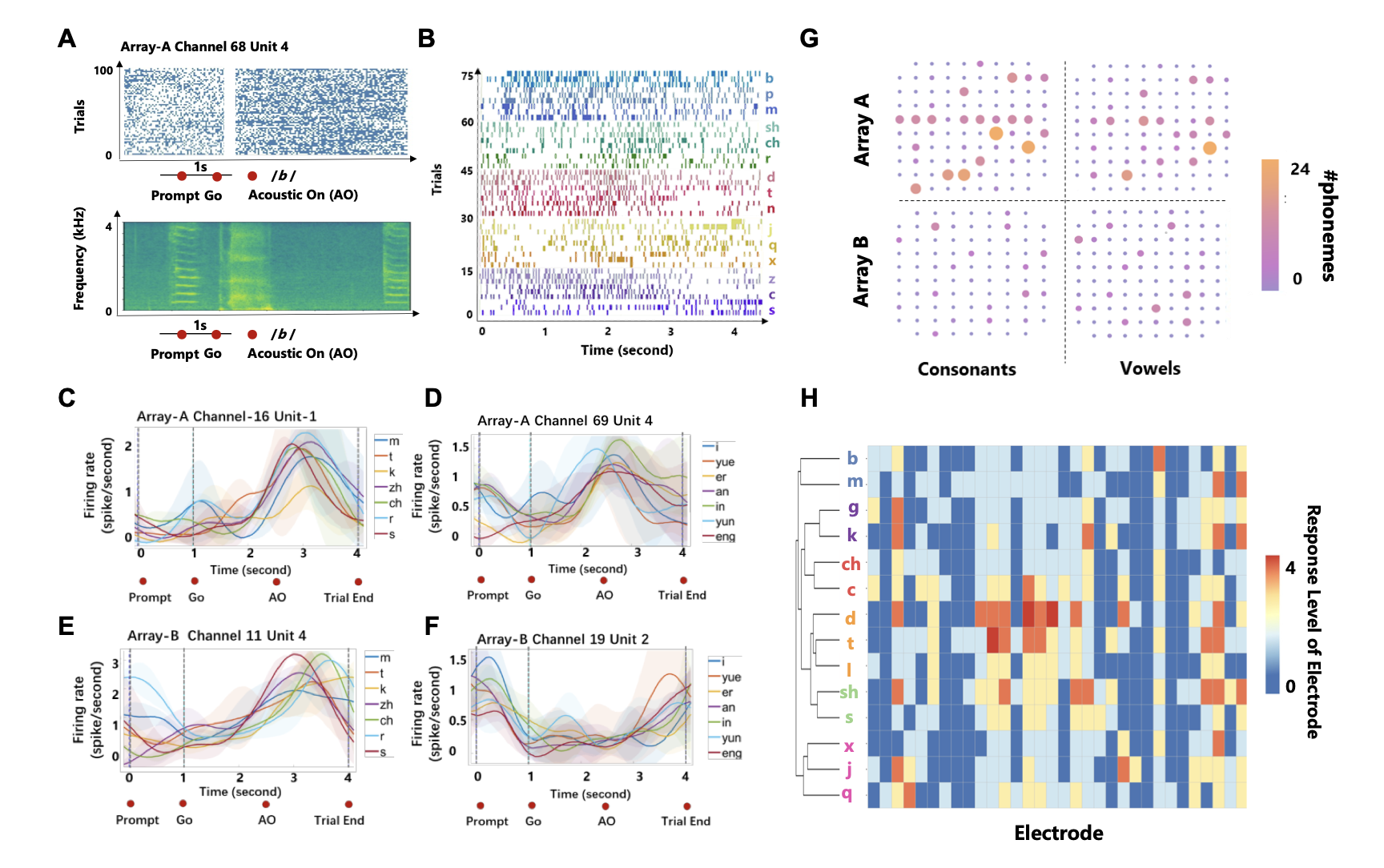}
  \caption{Analysis of neural activities during phoneme speaking. (A) Raster plot of an example neuron across repetitive trials of consonant $/b/$. The spectrogram shows an example of audio data. (B) Raster plot of an example neuron across repetitive trials of 15 different consonants. Consonants with similar articulations are grouped in similar colors. (C-F) Trial-averaged firing rates (mean ± s.d.) of single neurons during the speaking of consonants (C and E), and vowels (D and F). The three dashed lines indicate the timestamps of `Prompt', `Go', `AO', and `Trial End', respectively. (G) The electrode map of the significantly modulated channels for consonants (left column) and vowels (right column), respectively. The color and size of a channel indicate the number of consonants/vowels a channel significantly modulated. (H) Modulation matrix of electrodes. Each row refers to a consonant, and each column refers to an electrode. The left of the plot shows clustering results across electrode modulation.
}
  \label{fig:visualize}
\end{figure}

Figure \ref{fig:visualize}A (top) shows the raster plot of an example neuron across repetitive trials when the participant was prompted to pronounce the consonant $/b/$. The horizontal axis is aligned in time by the stage `Prompt' and stage `AO', respectively. Figure \ref{fig:visualize}A (bottom) is an audio spectrogram of consonants $/b/$ collected in a single trial.
Overall, stronger neural responses are observed during the phoneme speaking stage compared with the prompt stage, where the audio is played. The result is with significance (p \textless 0.01, sign-rank test) and is consistent with the previous studies \cite{stavisky2019neural}. Then we examine the neural activity during the speaking of different phonemes.
Figure \ref{fig:visualize}B shows the raster plot of an example neuron across repetitive trials of the participant speaking 15 different consonants. The results reflect the underlying consistency of neural activity across repetitive trials.

Figure \ref{fig:visualize}C-F plots the trial average firing rates of 4 example neurons. In each subfigure, the solid lines are the firing rates of neurons, and the standard deviations are indicated by the shaded areas. Three dashed lines in the subfigures indicate the timestamp of stage `Prompt', `Go', and `Trial End', while the stage `AO' occurs between 0.5s and 1s after stage `Go', denoted at the bottom of the figures. As shown in Figure \ref{fig:visualize}C-F, each curve with different color indicates a phoneme. The result shows that the tuning of spoken phonemes is intermixed with single units.

Figure \ref{fig:visualize}G analyzes the electrode distribution of the neural responses for different phonemes.
Overall, neural signals recorded with different electrodes show broad responses for phoneme speaking, and most of the electrodes modulate multiple phonemes. Figure \ref{fig:visualize}G plots the two 96-electrode arrays, and each circle indicates a single electrode that has a different size and color. The circle's color indicates the number of phonemes significantly modulated by the electrode (p \textless 0.05, sign-rank test).
As shown in Figure \ref{fig:visualize}G, active electrodes are distributed throughout the area sampled by the arrays, and most modulated to multiple phonemes, suggesting a distributed coding scheme.

\subsubsection{Analysis of phonemic similarity in neural representations}

In Figure \ref{fig:visualize}B, the consonants are divided into 5 groups according to \cite{duanmu2007phonology} (indicated by the colors of phonemes in Figure \ref{fig:visualize}B) by the types of articulator movement. As shown in Figure \ref{fig:visualize}B, the spike activities are significantly more similar between phonemes within the same group than between phonemes from different groups (p \textless 0.01, shuffle test). The results reflect the underlying consistency between the neural representations and articulator movements in phoneme speaking, namely, phonemes with similar articulations are also similar in neural signals.

To further evaluate the phonemic similarity in neural activity, we compute the hierarchical clustering results with the single-electrode response of different phonemes. Figure \ref{fig:visualize}H plots the single-electrode response for speaking 21 consonants. In the matrix, each column corresponds to a single electrode, and each row corresponds to a single consonant. Each matrix element $i,j$ indicates the modulation intensity of electrode $j$ to consonant $i$. The modulation intensity has four levels according to the firing rate. Only the electrodes with an intensity greater than 3 are plotted for visualization purposes. The left side of the figure shows the hierarchical clustering result with the single-electrode response across different consonants, where consonants with similar articulations are shown in the same color. Results indicate that the hierarchical clustering based on neural responses shows consistency with the articulation of phonemes, which further demonstrates that the neural representations and articulator movements in phoneme speaking share a similar structure.

\begin{figure*}[t]
  \centering
  \includegraphics[width=1\linewidth]{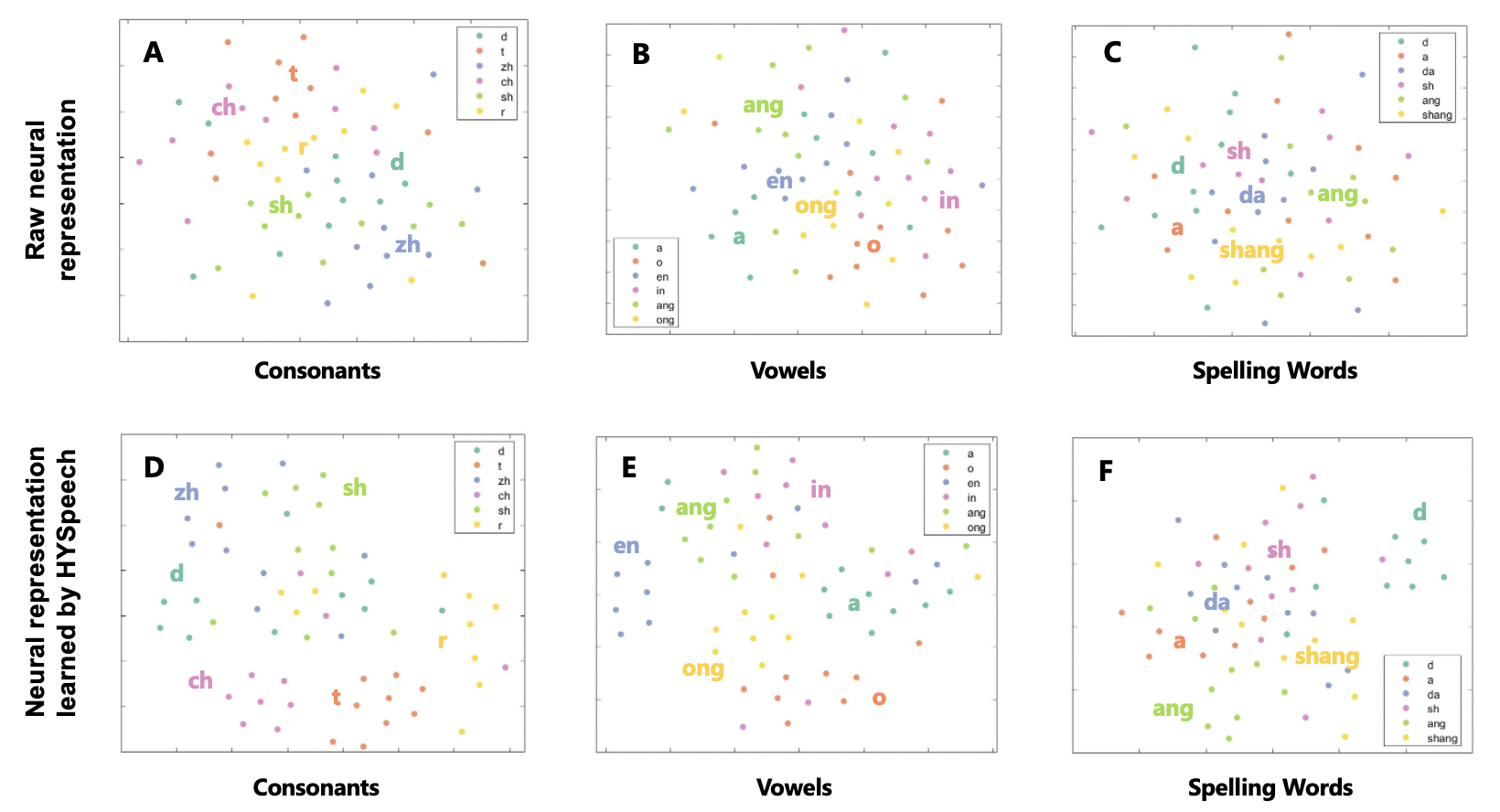}
  \caption{Visualization of neural representations. (A-C) $t$-SNE of raw neural representations of consonants (left), vowels (middle), and words (right), respectively. Each point corresponds to a single trial. (D-F) $t$-SNE of neural representations learned with HYSpeech.}
  \label{fig:3}
\end{figure*}%

\subsection{Neural phonemic structures learned by HYSpeech}
\subsubsection{Analysis of hyperbolic-based neural representation learning}

Firstly, we inspect the learned neural representation in comparison with the raw neural signals with $t$-SNE (Figure \ref{fig:3}A). Overall, with the raw neural signals, phonemes with similar articulations are closely spaced and difficult to discriminate from each other, similar to findings in \cite{moses2019real} and \cite{stavisky2019neural}, which may lead to confusion and errors in classification. While after hyperbolic-based learning, the neural representations are more discriminative in space. Especially, phonemes with similar articulations are well separated in space showing high discriminative ability.

For example, the consonant $/sh/$ and $/r/$ are similar in both articulator movement and manner of articulation (see Figure \ref{fig:new}A), so that the neural responses are close and confused in the raw space (Figure \ref{fig:3}A). While in the learned neural representation space, the representation of $/sh/$ and $/r/$ are distant from each other and well separated (Figure \ref{fig:3}D). Similar findings can be found with the vowels and words. In Figure \ref{fig:3}B, the similar vowels of $/o/$ and $/ong/$ are closely placed in the raw space, while they are well separated in the representation space (Figure \ref{fig:3}E). The word-spelling task further confirms the observation (Figure \ref{fig:3}C and F). The results demonstrate that, by taking the hierarchical structure of phoneme articulation as prior knowledge, the proposed HYSpeech approach learns to extend the space for similar phonemes such that they can be separated in space effectively.

\subsubsection{Analysis of the learned neural structures}

Here, we visualize the learned neural structure for phoneme speaking. Based on the aforementioned findings, there is a phonemic similarity in neural representations. Then the natural question is how the learned neural hierarchical structure reveals phoneme structure in neural representations.

\begin{figure*}[t]
  \centering
  \includegraphics[width=1\linewidth]{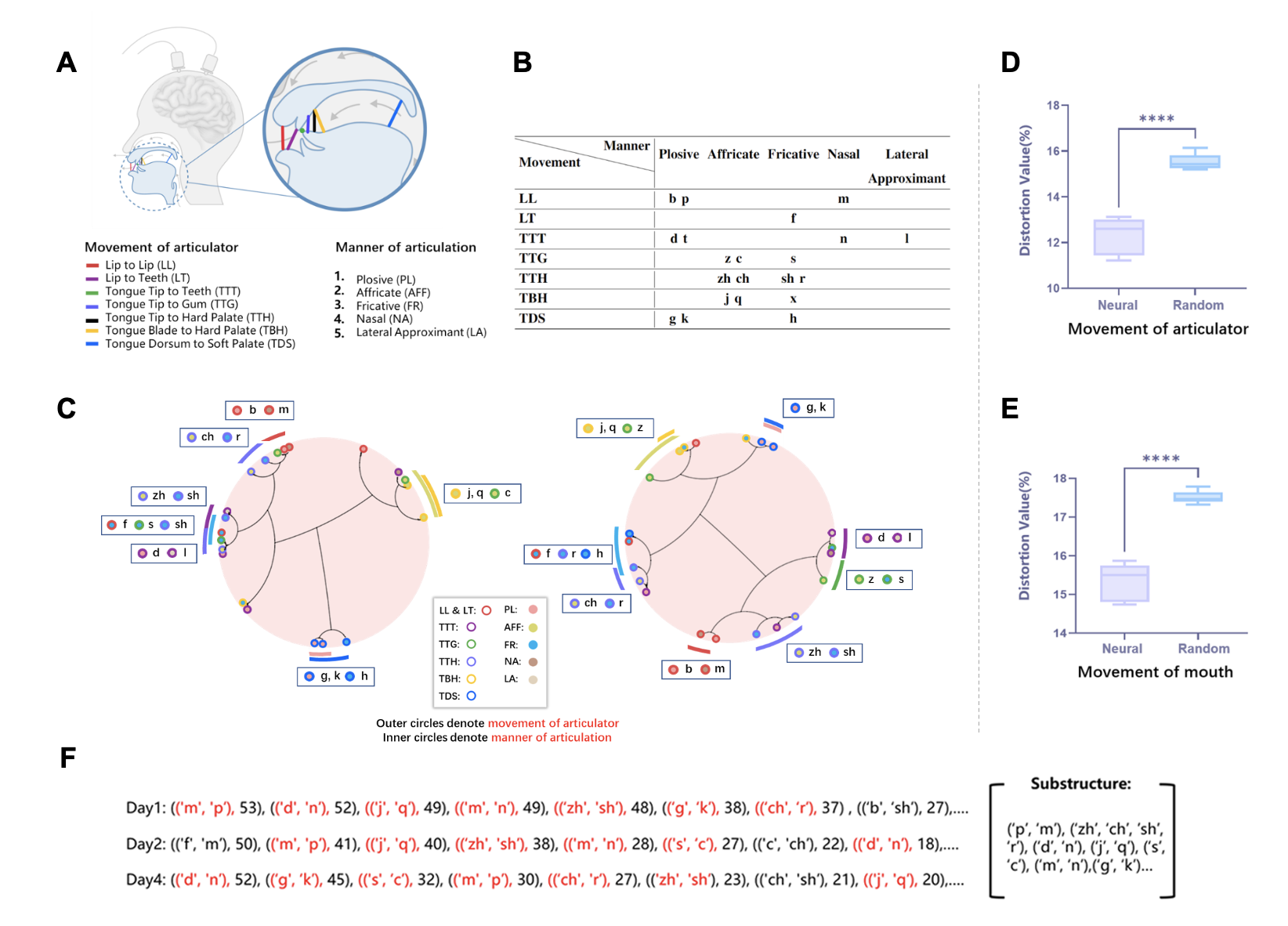}
  \caption{(A) Articulators' movements during the participant spoke different Mandarin consonants. (B) Categorization of Mandarin consonants according to the articulations.  (C) Visualization of the learned hierarchical clustering tree by HYSpeech from Day-1 and Day-2 consonant datasets. (D-E) The distortion value of the hierarchical clustering tree of consonants and vowels versus random levels, where a lower distortion value indicates better representations in space. (F) The learned common substructures in hierarchical clustering trees of across multiple consonant datasets and days. }
  \label{fig:new}
\end{figure*}

Taking the consonants as an example. Figure \ref{fig:new}A illustrates the articulations of different consonants, and Figure \ref{fig:new}B gives the categorization of consonants according to the articulations \cite{duanmu2007phonology}.
In Figure \ref{fig:new}C, we plot the hierarchical clustering tree of 21 consonants of two-day datasets learned by our model. In the tree-building process, the embeddings of trials calculated by our model are averaged according to their categories and projected as nodes to the Poincar$\acute{e}$ disk's edge and connected based on the distance between the nodes to form a sub-tree, and then a hierarchical clustering tree is built from the bottom up. We use different colors of the inner and outer circles to indicate different types of consonants. The color of the outer circles indicates different movements of the articulator, and the color of the inner circles indicates different manners of articulation (LF is grouped with LL because of its close position to LL, as shown in the legend of Figure \ref{fig:new}C). Outside the pink Poincar$\acute{e}$ disk, clusters of the same category are indicated by arcs of corresponding color, where the arcs indicating the movement of the articulator are on the outside and the arcs indicating the manner of articulation are on the inside.

Overall, the learned neural structure is mostly consistent with the articulation-based structure. For instance, at the bottom of the Day-1 clustering tree and the top right of the Day-2 clustering tree, $/g/$, and $/k/$ form a cluster. Both $/g/$ and $/k/$ are tongue dorsum and soft palate forming obstructions (blue outer circle), while both are plosive (the pink inner circle). The difference between the two is that $/k/$ is aspirated and $/g/$ is not. There are two main observations from these hierarchical clustering trees. First, the inner and outer circles with the same color are clustered more frequently in the hierarchical clustering tree, as shown in the box around the clustering tree. Second, there are multiple similar clusters in clustering trees of different days, e.g. $/b/$ and $/m/$, $/g/$, and $/k/$.

We further compare the learned structure from different days to evaluate the consistency over days. In Figure \ref{fig:new}F, we count the frequency of occurrence of substructures and find a series of high-frequency common substructures in the data of different days. First, we perform the clustering procedure 100 times on the dataset each day. Second, we count the number of binary clusters in hundred hierarchical clustering trees for different days. We then sort and pick out the high-frequency common binary clusters on different days, which are indicated in red font. Finally, we aggregate these high-frequency common clusters into a multi-day common substructure. The learned common substructure is similar to the articulation-based structure which reveals a phoneme structure.

\subsubsection{Effectiveness of hierarchical structures in learning}

Here we compare different settings of the hierarchical structure in the hyperbolic-based learning process, to demonstrate the effectiveness of hierarchical clustering.

\textit{1) Random v.s. learned structures.} We first compare randomly assigned hierarchical structures against the learned hyperbolic-based hierarchical structures from the neural signals.
To generate random hierarchical structures, we perform clustering with random noises instead of neural signals. We ran 100 times to obtain 100 random hierarchical structures.
For evaluation, we calculate the distortion value of these trees, which can be defined as \cite{bradley1998refining}:
\begin{equation}
    Distortion = \frac{\sum_i^n\sum_j^{n-1}{\mathbbm{1}(class(i) =  class(j))s_{ij}}}{\sum_i^n\sum_j^{n-1}{\mathbbm{1}(class(i) \ne class(j))s_{ij}}}
\end{equation}
where $s_{ij}$ is the shortest distance between nodes $i,j$ on the clustering tree and $class(\cdot)$ is the classification of phonemes. The lower this distortion value is, the more similar the clustering tree is to the classification structure of phonemes.

We compare the distortion of the neural-based structure and the random-based structure on consonants classification (by the movement of the articulator) and vowels classification (by the movement of the mouth), as shown in Figure A1C, D, respectively. The distortion of the neural-based structure is significantly lower than that of the random-based structure both on consonants and vowels (p \textless 0.0001, sign-rank test), which indicates the superiority and the effectiveness of the hyperbolic-based hierarchical structure learning from neural signals.

\textit{2) Using articulation-based structures v.s. learning the substructures.}
Now that the learned hierarchical structures from the neural signals show similarity to articulation-based structures, an interesting question is to what extent, the phoneme structure in neural representation is consistent with the articulation-based structure.

To investigate this, we first assume that the phoneme structure in the neural representation is identical to the articulation-based structure, then the complete articulation-based structure will be more consistent with the data prior than the learned common substructures. Therefore, we compare the articulation-based structure with the learned common substructure as a constraint in the structure learning process to constrain that the learned structure should be similar to the constrained structure. Specifically, on our model's loss, we add the distance constraint and remove the clustering loss. The distance constraint is based on a category appointed in advance, and the data in the same category are constrained to be closer together, which can be defined as:

\begin{equation}
    Constraint = \sum_i^n\sum_j^{n-1}{\mathbbm{1}(class(i) = class(j))} d_{ij}
\end{equation}
where $d_{ij}$ is the hyperbolic distance between $i,j$ and $class(\cdot)$ is the category of phonemes.
Comparison is conducted with three conditions: 1) constraint based on the articulation-based structure (see Figure A1C), 2) learned common substructure from neural signals (see Figure \ref{fig:new}F), and 3) without constraint.

The results are given in Table \ref{table2}. The learned common substructure achieves 56.54\%, 74.84\%, and 58.47\% accuracy, which outperforms the articulation-based structure by 55.27\%, 73.99\%, and 57.90\%. While both structures outperform the unconstrained condition by 52.37\%, 70.03\%, and 55.10\%, indicating that the phoneme structure in the neural representation is similar to, but not fully the same as, the articulation-based structure of phonemes.

\subsection{Phoneme classification performance and compraison}
Here we evaluate the phoneme classification performance of the proposed HYSpeech approach in comparison with existing methods. For each method, the number of neurons in the input layer is the dimension of the input Euclidean vector $x^E$ and the number of neurons in the output layer is the number of categories for the classification task. We test all methods on a four-day clinical dataset for three tasks (consonants, vowels, and words). Due to our small-scaled dataset, all the following accuracy rates are calculated by the leave-one-out method to ensure the validity of the performance test. We compute standard deviations of our results over 5 random runs.
The competitors are specified as follows:
\begin{itemize}[leftmargin = 15pt]
 \item \textbf{SVM.} An SVM with a linear kernel is adapted to provide a baseline performance.
  \item \textbf{GRU.} Considering that the neural signal is a time series, we use GRU as another baseline method. Here, we set the input layer dimension of GRU to the number of neurons and the output layer dimension to the number of classification categories. One hidden layer is employed and its dimension is set to 256. The parameters are optimized with the Adam algorithm with a learning rate of 0.05.
  \item \textbf{gHHC.} gHHC is a hierarchical clustering method on hyperbolic space proposed in \cite{monath2019gradient}. For comparison, we replace the clustering loss of our model with the loss function of gHHC, and the rest of the settings are consistent with our model.
  \item \textbf{HypHC.} HypHC is a hyperbolic hierarchical clustering approach proposed in \cite{chami2020trees}. HypHC notes that the internal structure can be directly inferred from the leaf nodes, thus directly optimizing the leaf node positions. Similarly, we replace the clustering loss of our model with the loss function of HypHC, and the rest of the settings are consistent with our model.
  \item \textbf{HMCN.} HMCN is a typical multi-label classification method proposed in \cite{wehrmann2018hierarchical}. Considering the multiple ways of classifying phonemes, which can carry multiple labels, we use HMCN to perform multi-label classification of phonemes (consonants and vowels), while not for the word task. The network structure consists of multiple local output layers (corresponding to each layer of the hierarchy) and a global output layer with a mixture of local and global information. The number of neurons in each hidden layer is 384. The global output layer computes cross-entropy loss with multi-hot labels, and the local output layer is one-hot labels at different levels.

 \begin{figure*}[t]
  \centering
  \includegraphics[width=1\linewidth]{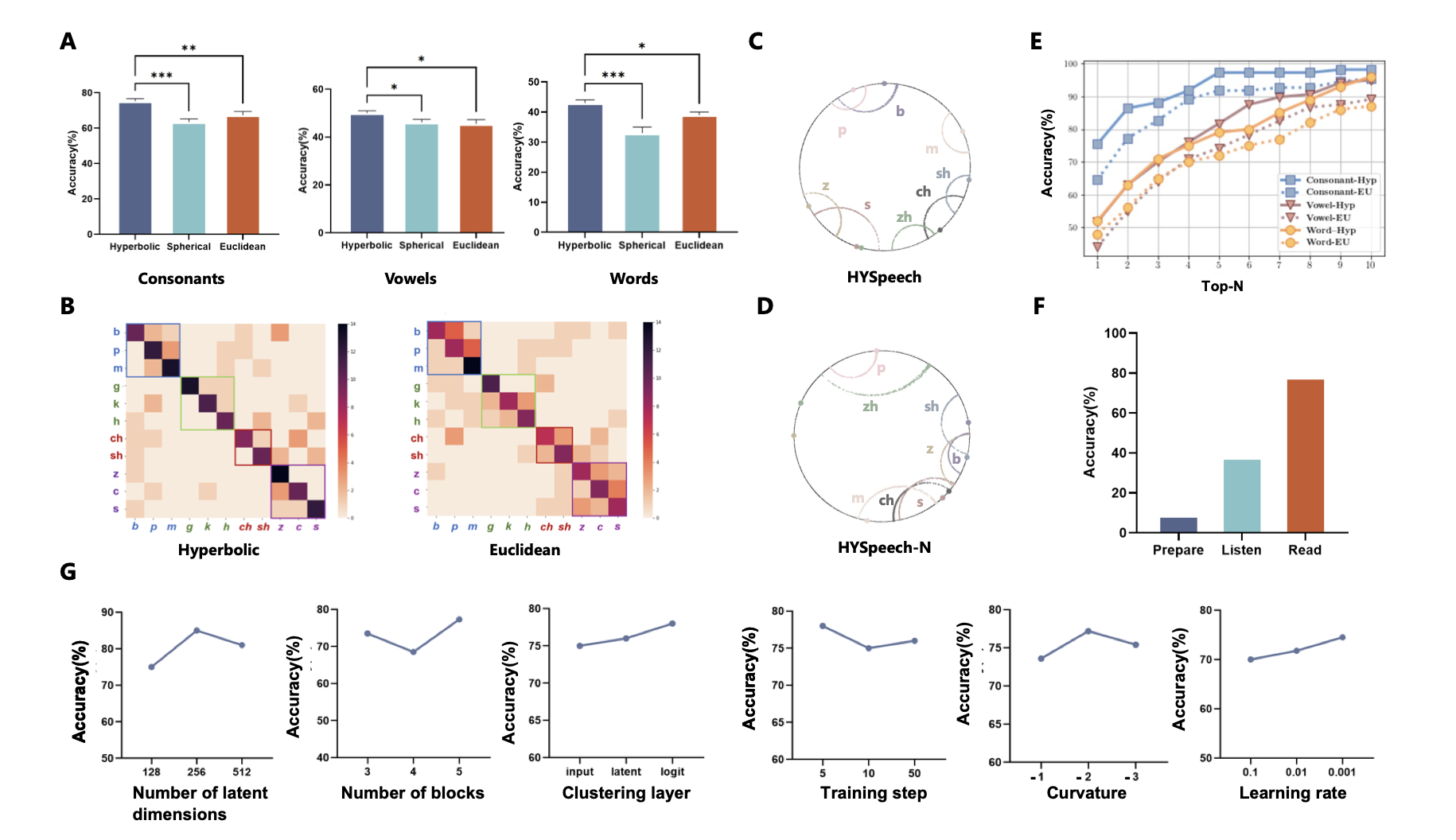}
  \caption{Performance and comparison. (A) Classification accuracy of 21 consonants, 24 vowels, and 20 words in different spaces. Significance levels: * - p \textless 0.05, ** - p \textless 0.01 and *** - p \textless 0.001. (B) Confusion matrices of consonant classification in different spaces, the colored rectangles indicate consonants with similar articulations. (C-D) The decision boundaries of HYSpeech and HYSpeech-N (without clustering loss). (E) Comparison of the Top-N accuracy between our approach and the Euclidean space-based approach. (F) Classification performance of neural signals using neural signals recorded from different conditions. (G) Influence of different parameters on classification performance. }
  \label{fig:4}
\end{figure*}%

\end{itemize}

\begin{table*}

  \caption{Classification accuracies (\%) of different phonemes.}
  \label{tb:main}
  \centering

  \resizebox{1\columnwidth}{!}{
  \begin{tabular}{l|cc|c|cc|ccc}

    \toprule
    \multirow{2}{*}{Model}  & \multicolumn{2}{c}{Day-1} & \multicolumn{1}{c}{Day-2} & \multicolumn{2}{c}{Day-3} & \multicolumn{3}{c}{Day-4}  \\
    \cmidrule(r){2-3} \cmidrule(r){4-4} \cmidrule(r){5-6} \cmidrule(r){7-9}  & Consonant & Vowel & Consonant  & Word-1 & Word-2 & Consonant & Word-1 & Word-2\\

    \midrule
        \textbf{SVM}& 51.82  & 45.83 &  64.55  & 54.00 & 46.00 & 56.19 & 54.00 & 44.00   \\
    \textbf{GRU}& $46.95 \pm 1.29$  & $42.40 \pm 1.64$ &  $59.45 \pm 1.99$  & $49.20 \pm 2.16$ & $43.80 \pm 1.10$ & $54.28 \pm 0.96$ & $41.00 \pm 1.41$ & $41.40 \pm 1.52$    \\

      \hline
    \midrule
     \textbf{gHHC \cite{monath2019gradient}}&  $49.45 \pm 1.77$ & $41.60 \pm 0.82$  & $56.32 \pm 0.72$ & $48.60 \pm 0.89$ & $41.80 \pm 0.45$ & $55.05 \pm 0.43$ & $42.20 \pm 1.30$ &  $39.00 \pm 0.71$  \\
     \textbf{HypHC \cite{chami2020trees}}& $51.09 \pm 0.41$  & $46.66 \pm 1.18$  & $66.07 \pm 0.85$ & $53.40 \pm 0.30$ & $46.80 \pm 1.79$ & $56.77 \pm 1.27$ & $47.20 \pm 1.64$ & $41.20 \pm 0.45$   \\
    \textbf{HMCN \cite{wehrmann2018hierarchical}}& $49.64 \pm 2.19$  &$43.50 \pm 1.37$  & $58.18 \pm 1.58$  & - & - & $54.66 \pm 0.53$& - & - \\

    \hline
    \midrule
    \textbf{HYSpeech-EU}
           &$51.54 \pm 0.89$  & $43.87 \pm 0.04$ & $64.18 \pm 1.80$  & $49.60 \pm 1.82$  &$42.20 \pm 1.64$  & $51.80 \pm 0.84$ &$47.40 \pm 1.67$  & $41.80 \pm 0.84$ \\
    \textbf{HYSpeech-N}
           & $52.37 \pm 0.82$  & $46.24 \pm 2.94$ &  $70.03 \pm 1.11$ &  $53.60 \pm 1.14$ & $44.80 \pm 1.64$ & $55.10 \pm 1.96$ & $50.20 \pm 1.92$ & $43.40 \pm 1.52$ \\
    \textbf{HYSpeech (ours)}
           & $\textbf{58.03} \pm \textbf{2.58} $   & $\textbf{51.25} \pm \textbf{0.02}$  &  $\textbf{75.21} \pm \textbf{1.43}$ &  $\textbf{57.00} \pm \textbf{1.41}$ & $\textbf{51.50} \pm \textbf{2.12}$ & $\textbf{61.42} \pm \textbf{2.02}$ & $\textbf{55.40} \pm \textbf{0.71}$ & $\textbf{48.80} \pm \textbf{1.92}$ \\
    \bottomrule
  \end{tabular}}
  \label{tab:class}
\end{table*}%

\begin{table}[t]
\caption{Model performance using different distance constraints.}
\centering
\resizebox{.75\columnwidth}{!}{
\begin{tabular}{l|ccc}
\toprule
 HYPSpeech & Day-1 & Day-2 & Day-4\\
\midrule
 \textbf{Articulation-based structure} & $55.27 \pm 0.99$ & $73.99 \pm 0.49$ & $57.90 \pm 1.04$\\
\midrule
\textbf{Without constraint} & $52.37 \pm 0.82$ & $70.03 \pm 1.11$ & $55.10 \pm 1.96$\\
\midrule
\textbf{Neural substructure} & $\textbf{56.54} \pm \textbf{0.40}$ & $\textbf{74.84} \pm \textbf{0.52}$ & $\textbf{58.47} \pm \textbf{1.27}$\\
\bottomrule
\end{tabular}}

\label{table2}
\end{table}

\subsubsection{Comparison with non-hierarchical baselines}
To compare the proposed HYSpeech with traditional methods of SVM and GRU, we evaluate the classification performance of different tasks on each day. Overall, HYSpeech outperforms SVM and GRU on all tasks. As shown in Table \ref{tb:main}, in the consonant task, HYSpeech achieves 58.03\% (Day-1), 75.21\% (Day-2), and 61.42\% (Day-4) accuracy, which is significantly higher than 46.95\%-59.45\% of GRU, and 51.82\%-64.55\% of SVM. With the vowel task, the performance of HYSpeech is 51.25\% (Day-1) accuracy, which is 8.85\% and 5.42\% higher than GRU and SVM, respectively. In the word task, the average classification accuracies of two-word sets of HYSpeech are 54.25\% (Day-3) and 52.10\% (Day-4), which outperform GRU by 46.50\%, 41.20\%, and SVM by 50.00\% and 49.00\%, respectively.

\subsubsection{Comparison with hierarchical-based approaches}
We further compare the classification performance with hierarchical methods. As shown in Table \ref{tb:main}, overall, our method achieves the best performance on all tasks. Specifically, HYSpeech outperforms gHHC, HypHC, and HMCN by about 46.75\%, 51.15\%, and 51.49\%, respectively, demonstrating the strength of the proposed structure and the hierarchical clustering process.

\subsubsection{Evaluation of model variants}
Here we use three different variants of our model to examine the effectiveness of our method, and the results are shown in Table \ref{tb:main}. The model HYSpeech-EU is constructed by transforming all operations of our model to Euclidean space. Our approach outperforms HYSpeech-EU with all the tasks, suggesting the essentiality of using the hyperbolic space. For HYSpeech-N, we remove clustering loss, which obtains decreased performance compared with our approach, indicating the importance of the clustering part.

\subsection{Model evaluation}
 Here we evaluate the performance of the proposed approach under different settings, including the effectiveness of model components and performance under different parameter settings.

\subsubsection{Phoneme classification in different spaces}

Here we analyze the phoneme classification performance with different spaces. Figure \ref{fig:4}A plots the classification accuracy of consonants, vowels, and words in three spaces (hyperbolic, spherical, and Euclidean), respectively. The classification performance in hyperbolic significantly outperforms the performance in the spherical and Euclidean spaces for all three tasks. We further compare the classification confusion matrix of hyperbolic and Euclidean spaces (consonant task) in Figure \ref{fig:4}B. With the Euclidean space, phonemes with similar articulations are confused, which decreases the classification performance. With the hyperbolic space, the phonemes are clearly separated, indicating that the phoneme representations are the most discriminative in hyperbolic space.

Figure \ref{fig:4}E presents the Top-N performance of the proposed approach, which refers to the accuracy that one of the first N answers given by the model is correct. The proposed approach obtains a Top-5 accuracy of 97.27\%, 81.67\%, and 79\% with the consonant, vowel, and word tasks, respectively. The Top-10 performance are 98.18\%, 95\%, and 96\%.

\subsubsection{Effectiveness of the hierarchical clustering constraint}
Then we compare the decision boundaries with and without clustering loss.
 Figure \ref{fig:4}C-D plots the decision boundaries of HYSpeech and HYSpeech-N. Owing to the hyperbolic MLR used in our model, the decision boundary is a curve. For visualization purposes, we set our latent dimension to 2. In Figure \ref{fig:4}C-D, the circles indicate the Poincar$\acute{e}$ disk, and the arcs of different colors in the circles indicate the decision boundaries of different consonants. Results show that the decision boundaries of HYSpeech are more clearly spaced with our clustering loss, demonstrating the effectiveness and strength of the clustering loss in learning discriminative representations.

 \subsubsection{Performance of neural signals recorded in different states}
We analyze the classification performance using neural signals recorded from different conditions of `prepare', `listen', and `read'. The experiment is conducted using the consonant datasets. Specifically, the stage `prepare' refers to the range from one second before 'Prompt' to 'Prompt', the stage `listen' refers to the range from 'Prompt' to 'Go', and the stage `read' refers to the range from 'AO' to one second after 'AO'. As shown in Figure \ref{fig:4}F, the classification performance is close to random during the preparation phase. While in the listening phase, the classification accuracy is better than random. Overall, the highest performance is obtained with the `read' phase.

\subsubsection{Evaluation with different model settings}
Here we analyze the effect of hyperparameters of our model. The experiments are conducted using consonant datasets. We analyze the effect of the latent dimension of the hyperbolic network on the classification performance. As shown in Figure \ref{fig:4}G, we compare three dimensions and achieve the best performance at 256. The effect of different learning rates on classification performance is also evaluated. The optimal performance is obtained with a learning rate of 0.001. The parameter of the number of blocks refers to the amount of data used in training and testing. Each block contains a set of data points of all categories. As shown in Figure \ref{fig:4}G, the optimal performance is given at 5 blocks, and the performance may be further improved if more data can be collected. The parameter of curvature (namely $c$ in Equation 2) indicates how curved a hyperbolic space is. We compare the effect of curvature on classification performance in hyperbolic space. We compare three curvatures and achieve the best performance at -2. Then we compare applying the clustering on the input, latent, and logit layers and the best performance is achieved when clustering is performed on the logit layer. Two training methods of joint training and alternating training, are also evaluated, for the hyperparameters of $\lambda$ and $\gamma$. In the joint training, we initialize $\lambda = 0$ and $\gamma = 1$, and after 100 epochs, we adjust them to an equal ratio of 0.5. In alternating training, we fix the number of iteration steps $k$, and let $\lambda$ and $\gamma$ switch directly between 0 and 1 every $k$ steps. Figure \ref{fig:4}G shows the classification performance at different steps and reaches the best performance at 5 steps.

\section{Conclusion}
In this work, we propose a hyperbolic model to decode spoken Chinese phonemes from neural signals. Our approach obtained superior performance compared with existing methods and achieved state-of-the-art. The significant performance improvement demonstrates that the neural representation of spoken phonemes contains a hierarchical structure, and using hyperbolic space for computation can be a suitable way to deal with the problem and can potentially bring further developments to the area. The findings suggest the feasibility of constructing high-performance Chinese speech BCIs based on phoneme decoding. The proposed idea and methodology are also beneficial for a broad area of neural decoding research.

\Acknowledgements{This work was partly supported by grants from the National Key R\&D Program of
China (2018YFA0701400), the Key R\&D Program of Zhejiang (2022C03011), the Natural Science Foundation of China (62276228), the Fundamental Research Funds for the Central Universities, the Starry Night Science Fund of Zhejiang University Shanghai Institute for Advanced Study (SN-ZJU-SIAS-002), and the Lingang Laboratory (LG-QS-202202-04).}

\Supplements{Appendix A.}



\end{document}


\ArticleType{RESEARCH PAPER}
\Year{2019}
\Month{}
\Vol{}
\No{}
\DOI{}
\ArtNo{}
\ReceiveDate{}
\ReviseDate{}
\AcceptDate{}
\OnlineDate{}


\begin{appendix}

\section{Categorization of Mandarin phonemes according to the articulations}

\begin{figure}[ht]
  \centering
  \includegraphics[width=1\linewidth]{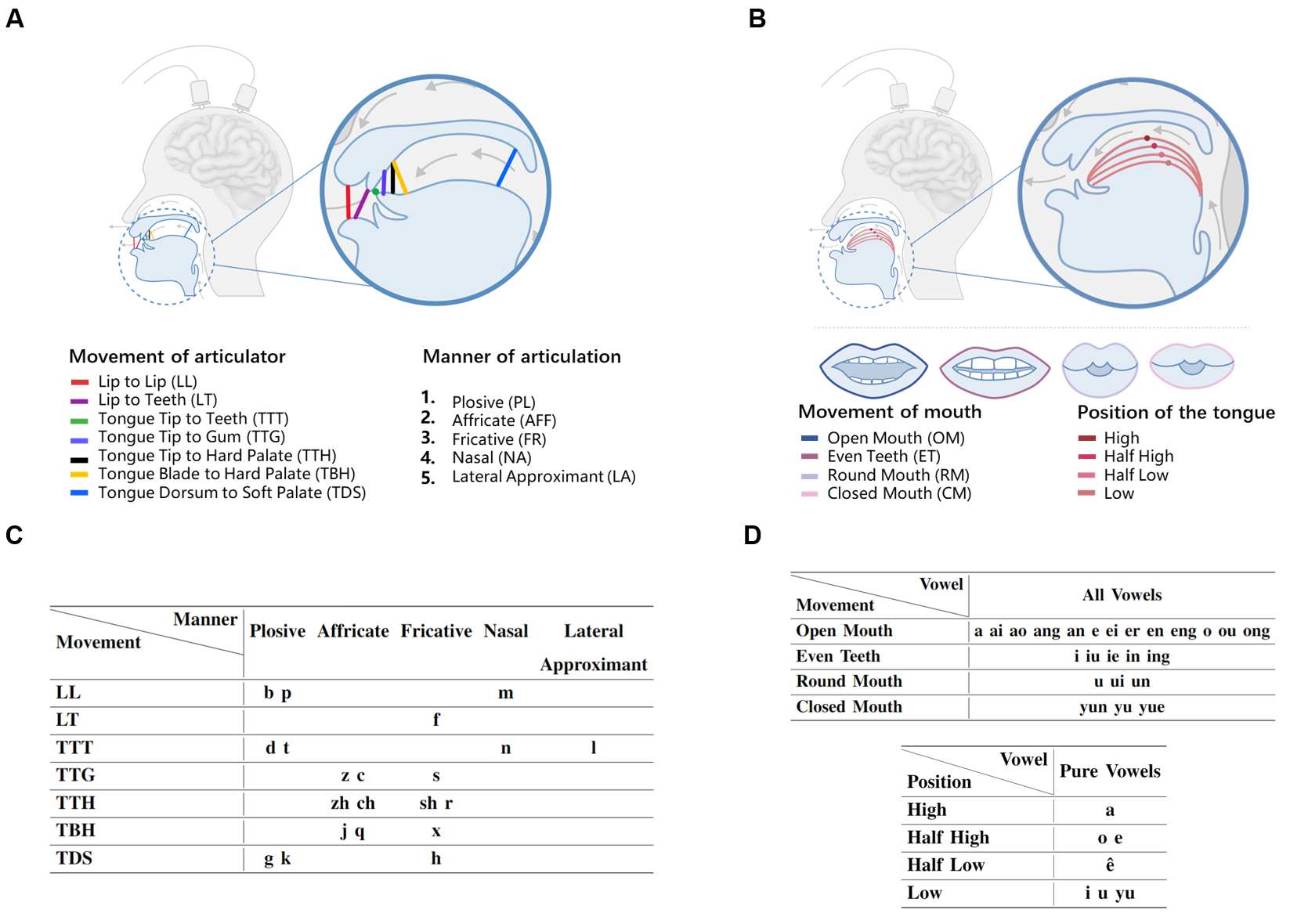}
  \caption{Categorization of Mandarin phonemes. (A-B) Articulators' movements during the participant spoke different Mandarin consonants and vowels. (C-D) Categorization of Mandarin consonants and vowels by the ways of pronunciations.}
  \label{fig:6}
\end{figure}

The Mandarin phonemes can be categorized according to the articulations \cite{duanmu2007phonology}. The basic places of articulation include the lip, teeth, tongue, gum, and palate, and phonemes’ pronunciation results from the sequential combination of the articulations.
We show the articulators’ movements during the speaking of consonants and vowels, respectively, due to their different way of pronunciations (see Figure \ref{fig:6}).

\begin{itemize}[leftmargin = 15pt]
\item
\textbf{Consonants.}
During the pronunciation of consonants, the articulators form an obstruction. There are two ways of categorizing consonants, according to \cite{duanmu2007phonology}.
From the perspective of movement of the articulator, the basic articulation of consonant pronunciation includes lip-to-lip (LL), lip-to-teeth (LT), tongue-tip-to-teeth (TTT), tongue-tip-to-gum (TTG), tongue-tip-to-hard-palate (TTH), tongue-blade-to-hard-palate (TBH), tongue-dorsum-to-soft-palate (TDS) (as shown with the colored lines in Figure \ref{fig:6}A). Taking the consonant $/b/$ as an example. $/b/$ is produced as a plosive by forming an obstruction through the contact of the upper lip with the lower lip, so we categorize the $/b/$ in lip-to-lip, as shown by the red line in Figure \ref{fig:6}A. From the perspective of the manner of articulation, the Mandarin phonemes can be divided into five types: plosive (PL), affricate (AFF), fricative (FR), nasal (NA), and lateral approximant (LA).

\item
\textbf{Vowels.}
The sound of vowels is mainly determined by the movement of the tongue and mouth, so we generally use the position of the tongue and movement of the mouth to classify vowels, according to \cite{duanmu2007phonology}. The tongue positions are shown in Figure \ref{fig:6}B with the red dots and arcs on the tongue. The dots indicate the front and back of the tongue, and the arcs indicate the height of the tongue. The movement of the mouth can be divided into four types: open mouth (OM), even mouth (ET), round mouth (RM), and closed mouth (CM), which are shown in Figure \ref{fig:new}B with four different colored lips.
\end{itemize}

A detailed classification of consonants and vowels can be seen in Figure \ref{fig:6}C-D. The phonemes in Mandarin can be considered a simplified version of the International Phonetic Alphabet (IPA) \cite{international1999handbook}. The consonants in Mandarin can be classified similarly to IPA, while it has fewer consonants and only contains pulmonic consonants.

\end{appendix}